\documentclass[conference]{IEEEtran}

\IEEEoverridecommandlockouts

\usepackage{cite}
\usepackage{amsmath,amssymb,amsfonts}
\usepackage{algorithm}
\usepackage{algpseudocode}
\usepackage{graphicx}
\usepackage{textcomp}
\usepackage{xcolor}
\usepackage{colortbl}
\usepackage{xurl}
\usepackage[hidelinks]{hyperref}
\usepackage{physics}
\usepackage{subcaption}
\usepackage{booktabs}
\usepackage{multirow}
\usepackage{enumitem}
\setlist[enumerate]{itemsep=5pt, topsep=5pt}
\setlist[itemize]{itemsep=5pt, topsep=5pt}

\usepackage{booktabs}
\usepackage{tabularx}
\usepackage{longtable}
\usepackage{array}
\usepackage{graphicx}
\usepackage{threeparttable}
\usepackage[most]{tcolorbox}
\usepackage{placeins}

\definecolor{resultbest}{HTML}{D9EAF7}
\definecolor{resultworst}{HTML}{FADBD8}
\definecolor{modelone}{HTML}{E97132}
\definecolor{modeltwo}{HTML}{4EA72E}
\definecolor{modelthree}{HTML}{0F9ED5}
\newcommand{\best}[1]{\cellcolor{resultbest}\textbf{#1}}
\newcommand{\worst}[1]{\cellcolor{resultworst}\textbf{#1}}
\newcommand{\voteone}{\cellcolor{modelone}Model 1}
\newcommand{\votetwo}{\cellcolor{modeltwo}Model 2}
\newcommand{\votethree}{\cellcolor{modelthree}Model 3}
\usepackage{fancyhdr}
\def\BibTeX{{\rm B\kern-.05em{\sc i\kern-.025em b}\kern-.08em
T\kern-.1667em\lower.7ex\hbox{E}\kern-.125emX}}

\makeatletter
\renewcommand{\IEEEauthorrefmark}[1]{\textsuperscript{#1}}
\makeatother

\begin{document}

\title{Autonomous Code Migration for Quantum Programming Languages: A Case Study with QED-C Benchmarks}

\author{
\IEEEauthorblockN{
    W. Michael Brown\IEEEauthorrefmark{1},
    Thien Nguyen\IEEEauthorrefmark{1},
    Avimita Chatterjee\IEEEauthorrefmark{2},
    Thomas Lubinski\IEEEauthorrefmark{3,4,5}
}
\IEEEauthorblockA{\IEEEauthorrefmark{1}\textit{NVIDIA, Santa Clara, CA 95051, USA}}
\IEEEauthorblockA{\IEEEauthorrefmark{2}\textit{Lawrence Berkeley National Laboratory, Berkeley, CA 94720, USA}}
\IEEEauthorblockA{\IEEEauthorrefmark{3}\textit{Quantum Computing Data, San Rafael, CA 94901, USA}}
\IEEEauthorblockA{\IEEEauthorrefmark{4}\textit{Cascade Quantum, San Rafael, CA 94901, USA}}
\IEEEauthorblockA{\IEEEauthorrefmark{5}\textit{QED-C Technical Advisory Committee -- Standards, Arlington, VA 22209, USA}}
}

\hypersetup{
  pdftitle={Autonomous Code Migration for Quantum Programming Languages: A Case Study with QED-C Benchmarks},
  pdfauthor={W. Michael Brown, Thien Nguyen, Avimita Chatterjee, Thomas Lubinski}
}

\maketitle

\begin{abstract}
Quantum programming requires concepts and intuition that differ substantially from those used in traditional programming languages. Multiple quantum programming frameworks address these requirements through different abstractions and execution models, creating tradeoffs in programmability, portability, and performance. Because quantum software is less prevalent in the training data available to large language models, autonomous coding agents might be expected to face additional challenges when working with these frameworks. We present a case study in autonomous migration of the QED-C Application-Oriented Benchmarks from Qiskit to CUDA-Q. Opus, GPT, and Gemini models were each tasked with porting thirteen benchmark methods spanning foundational algorithms, amplitude estimation, Monte Carlo methods, variational optimization, linear-system solving, and factoring. In all cases, the models were able to generate satisfactory ports with little direct coding help and there was a drastic reduction in the time required compared to a human engineer. However, in all cases, the agents resorted to shortcuts when implementation challenges surfaced and multiple prompts were required to force appropriate coding directions. Assessment of the generated implementations showed that no single model consistently produced the best port across all benchmarks, indicating that multi-agent ensemble strategies could mitigate challenges associated with quantum software. The effort resulted in publicly available, near-complete CUDA-Q support in the QED-C benchmarks along with model-generated skills designed to improve agent performance when implementing CUDA-Q applications. When tested on an unseen benchmark, these skills reduced the required model turns by up to 85\% and token usage by up to 96\%. We conclude by distilling lessons from the case study into suggested best practices for future agent-assisted development and migration of quantum software.

\end{abstract}
\vspace{1em}



\pagestyle{fancy}
\renewcommand{\headrulewidth}{0.0pt}
\lhead{}
\rhead{\thepage}
\renewcommand{\footrulewidth}{0.4pt}
\cfoot{}
\lfoot{Autonomous Code Migration for Quantum Programming Languages: A Case Study}
\rfoot{}

\section{Introduction}
\label{sec:introduction}

Quantum applications are commonly developed using high-level programming frameworks that combine circuit construction, classical orchestration, simulation, and access to quantum hardware. Prominent examples include IBM's Qiskit~\cite{JavadiAbhari2024}, Google's Cirq~\cite{Cirq2025}, Xanadu's PennyLane~\cite{Bergholm2018}, and NVIDIA's CUDA-Q~\cite{Kim2023}. These frameworks can emphasize different goals: Qiskit provides a broad circuit and algorithm ecosystem with close integration to IBM hardware, Cirq exposes detailed control for circuit construction and execution, PennyLane centers differentiable hybrid quantum--classical workflows, and CUDA-Q provides a C++ and Python programming model designed for heterogeneous CPU, GPU, and QPU computation~\cite{Heim2020}. All provide a path for GPU-accelerated simulation built on the NVIDIA cuQuantum libraries~\cite{Bayraktar2023,NVIDIA2026b}. Their differences in abstractions, supported operations, backend interfaces, differentiation facilities, and execution models nevertheless create consequential tradeoffs for portability, programmability, and performance. Automated migration can help developers navigate these tradeoffs by preserving an application's algorithmic intent while adapting its implementation to exploit the capabilities of another framework.

While the performance of automated code migration with LLMs can be very impressive for general programming languages~\cite{Pan2024,Zhang2025}, there are reasons to expect more challenges with quantum programming languages. Public code corpora used to train code models are highly uneven across programming languages~\cite{Kocetkov2023}. Quantum programming languages and frameworks are likely to be underrepresented relative to mainstream languages because they serve a comparatively small developer community, contain fewer public repositories and worked examples, and evolve rapidly as hardware capabilities and software abstractions change. Multilingual evaluations have shown the potential impact of language exposure. MultiPL-E found that model performance was correlated with programming-language popularity~\cite{Cassano2023}. A subsequent study found that performance generally increased with a language's proportion of the training corpus and that supplementing low-resource languages with validated training examples improved code generation~\cite{Cassano2024}.

In this work, we study autonomous quantum-software migration using general-purpose models. To evaluate whether these models can work across a range of quantum concepts, we sought a high-quality collection of cross-application algorithms implemented within a consistent software framework and accompanied by automated correctness checks and validation. The QED-C Application-Oriented Performance Benchmarks fit these requirements: the suite spans foundational algorithms, amplitude estimation, Monte Carlo methods, variational optimization, linear-system solving, and factoring; provides common execution and analysis infrastructure; and evaluates results using algorithm-appropriate measures of solution quality~\cite{QEDC2019,Lubinski2023}. We therefore tasked the agents with migrating thirteen benchmark methods from Qiskit to CUDA-Q while preserving the higher-level quantum algorithms.

We evaluated three models: Anthropic's Claude Opus~\cite{Anthropic2026b}, an OpenAI GPT model~\cite{OpenAI2026b}, and Google's Gemini~\cite{Google2026}. Because agentic coding is stochastic and highly path dependent, with small variations potentially producing divergent implementation trajectories, and because both models and agent harnesses are developing rapidly, we present these experiments as a case study rather than a controlled comparison. We therefore anonymize the models in the results. Our intent is not to rigorously rank individual models, but to characterize their general performance and extract practical lessons from autonomous migration of a substantial quantum-software codebase.

We evaluated the independently generated ports for functional correctness, migration effort, and differences in implementation structure. We then used cross-agent assessment to identify complementary strengths and support consolidation of the final implementations. Finally, we evaluated whether reusable knowledge distilled from the migrations could improve agent performance on unseen algorithms and transfer across models.

To our knowledge, this work is the first report of autonomous agent-based migration of a substantial application-oriented quantum benchmark suite directly between high-level quantum programming frameworks, together with an empirical analysis of migration effort, implementation changes, agent-ensemble consolidation, and reusable migration knowledge.

\begin{figure*}
    \centering
    \includegraphics[width=0.7\linewidth]{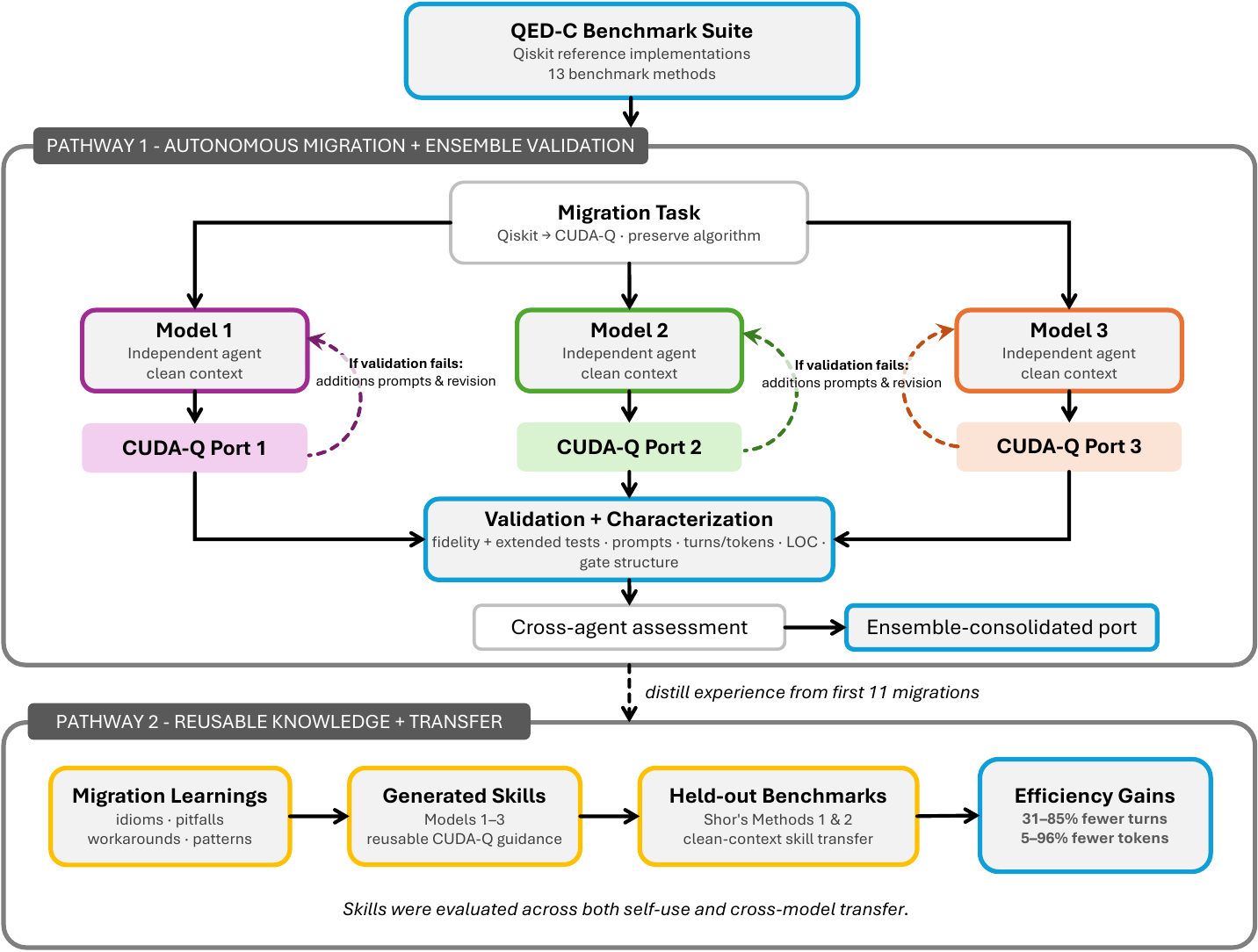}
   \caption{\textbf{Overview of the autonomous code-migration and knowledge-transfer methodology.}
   In Pathway 1, three coding agents independently migrate the 13 QED-C benchmark methods from Qiskit to CUDA-Q under isolated contexts. The resulting implementations are validated and characterized using fidelity, migration-effort, code-size, and gate-structure measures, followed by cross-agent assessment and ensemble consolidation. In Pathway 2, experience from the first 11 benchmark migrations is distilled into model-generated skills and evaluated on the held-out Shor’s Methods 1 and 2 under clean contexts, including both self-use and cross-model skill transfer. Skill use reduces model turns by 31–85\% and token usage by 5–96\% relative to the corresponding no-skill baselines.}
\label{fig:pipeline}
\end{figure*}


\section{Background} \label{sec:background}

\subsection{QED-C Application-Oriented Benchmarks}

The Quantum Economic Development Consortium (QED-C) Application-Oriented Performance Benchmarks provide an open-source suite for evaluating quantum computing systems using complete algorithms and small applications rather than isolated device-level operations~\cite{QEDC2019,Lubinski2024}. Developed under the QED-C Standards and Performance Metrics Technical Advisory Committee, the benchmarks characterize quantum computing performance from an application and end-user perspective. Each benchmark executes an algorithm over a range of problem sizes or input configurations, comparing results with analytic or classical references using algorithm-appropriate measures of solution quality.

This application-oriented approach complements lower-level characterization metrics by exercising the quantum computing stack, including compilation, circuit optimization, execution, and classical post-processing. The original QED-C methodology uses volumetric analysis to relate result fidelity to circuit width and depth, together with execution-time measures that reflect the quality and time-to-solution experienced by users~\cite{Lubinski2024}. Analogous to the SPEC benchmark methodology for classical systems, this approach enables systematic comparison of quantum computing platforms across architectures and problem types.

These properties also make the QED-C suite useful for evaluating quantum software portability. Its benchmarks span multiple algorithmic and programming patterns, while their common structure provides executable reference implementations and algorithm-specific correctness measures. Consequently, migration between programming frameworks can be evaluated using realistic quantum programs rather than isolated code-generation examples, while retaining a consistent mechanism for checking whether the migrated implementation reproduces the expected algorithmic behavior.

\subsection{Qiskit and CUDA-Q Programming Models}

Qiskit and CUDA-Q both support the development and execution of gate-based quantum algorithms, but they expose different programming abstractions. Qiskit centers programming around the flexible circuit builder pattern, via the \texttt{QuantumCircuit} abstraction, through which quantum and classical registers, operations, measurements, and control flow are explicitly assembled into circuit representations. These circuits can subsequently be transformed through Qiskit’s transpilation infrastructure, which performs operations such as synthesis, optimization, gate decomposition, and mapping to the constraints of an execution target~\cite{JavadiAbhari2024}. This separation between circuit construction and target-specific transformation provides a flexible representation that can be inspected and manipulated throughout the software workflow. 

CUDA-Q instead provides a single-source programming model in C++ and Python for accelerated quantum supercomputers designed to integrate QPUs with conventional CPU and GPU computation~\cite{Kim2023}. Its central abstraction is the quantum kernel, a callable containing computation intended for execution on a QPU or simulator. Quantum kernels generalize conventional circuit representations: they can be parameterized and composed from other kernels and can contain classical control flow, measurement-dependent conditionals, and classical computation~\cite{NVIDIA2026a}. CUDA-Q additionally provides language-level mechanisms for applying controlled and adjoint forms of kernels and operations, supporting the composition of larger quantum programs from reusable computational units. In CUDA-Q, quantum execution is facilitated by launcher primitives, such as sample, run, or observe. Depending on the circuit structure, e.g., with or without conditional qubit measurement feedback or algorithmic intents (getting raw measurements or estimating the expectation values), the appropriate launcher primitive should be used.

These differences make migration from Qiskit to CUDA-Q more than a syntactic translation between gate APIs. An algorithm expressed through Qiskit circuit construction may require restructuring when represented using CUDA-Q kernels, along with selection of execution primitives, particularly for dynamic control, mid-circuit measurement and classical feed-forward, controlled or adjoint subroutines, and parameterized computation. Moreover, the two frameworks can lower equivalent high-level operations differently during compilation, so similarity at the algorithmic level does not necessarily imply identical circuit structure or gate counts after decomposition. These programming-model differences motivate treating framework migration as a problem of preserving algorithmic intent while adapting the implementation to the abstractions and execution semantics of the target framework.

\subsection{Agentic Code Migration and Reusable Skills}

Large language models are increasingly used not only to generate isolated code fragments but also as software-engineering agents that interact with development environments to complete multi-step programming tasks. In this setting, an agent can inspect a repository, navigate and modify files, execute programs and tests, observe failures, and iteratively revise its implementation. SWE-bench formalized repository-level evaluation using real software-engineering issues that require changes to existing codebases rather than standalone program synthesis~\cite{Jimenez2023}. SWE-agent further demonstrated how an agent-computer interface can support autonomous repository navigation, file editing, command execution, and test-driven iteration~\cite{Yang2024}.

This agentic workflow is particularly relevant to code migration, where successful translation generally requires more than replacing individual language constructs. An agent must interpret the source implementation, identify corresponding abstractions in the target framework, modify code within an existing software architecture, and use execution feedback to identify and correct behavioral differences. The repository-level and execution-aware workflows studied in SWE-bench and SWE-agent illustrate the importance of environmental feedback and iterative modification for this class of software-engineering task~\cite{Jimenez2023,Yang2024}. In quantum software, such feedback is especially important because syntactically valid code may execute successfully while still differing from the source implementation in algorithmic structure or output behavior.

Repeated migrations also create an opportunity to retain reusable knowledge rather than requiring an agent to rediscover the same framework-specific solutions for each task. Prior work has explored external skill libraries as a mechanism for accumulating and reusing agent experience. Voyager, for example, incrementally constructs a library of executable skills and retrieves and composes those skills when solving new tasks, providing an early demonstration of persistent skill acquisition by an LLM-based agent without fine-tuning the underlying model~\cite{Wang2023}. More generally, prompting research has demonstrated that information supplied at inference time can substantially affect model behavior~\cite{White2023,Leidinger2023,Sclar2024}.

\subsection{LLMs for Quantum Programming}

Research on LLM support for quantum programming has progressed from code completion and small-circuit synthesis toward execution-aware generation, optimization, and repair. Dupuis et al.\ specialized a code model for Qiskit and introduced the execution-based Qiskit HumanEval benchmark~\cite{Dupuis2024,Vishwakarma2024} and Henderson et al.\ have evaluated an uncustomized GPT-4 on circuits~\cite{Henderson2025}. Subsequent benchmarks broadened both tasks and validation: QuanBench evaluates 44 tasks using functional correctness and process fidelity~\cite{Guo2025}; QHackBench uses PennyLane hackathon problems and multi-agent refinement~\cite{Basit2025}; QCoder incorporates simulator-derived hardware constraints and demonstrates the benefit of iterative feedback~\cite{Mikuriya2025}; and QuanBench+ extends evaluation and feedback-driven repair across Qiskit, PennyLane, and Cirq~\cite{Slim2026}. Beyond generating code from prompts, Agent-Q fine-tunes an LLM using more than 14,000 optimized QAOA, VQE, and adaptive-VQE circuits to generate parameterized OpenQASM~3 circuits and useful initial parameters~\cite{Jern2025}, while Campbell et al.\ use specialized agents, semantic analysis, and quantum error-correction feedback to iteratively improve generated programs~\cite{Campbell2025}. Debugging studies likewise progressed from ChatGPT repairing 29 of 38 Bugs4Q defects when supplied with sufficiently detailed failure information~\cite{Guo2024} to QBugLM, an agentic pipeline for bug injection, detection, repair, and simulation-based validation in OpenQASM~3.

In terms of porting code between frameworks, evidence from quantum-software migration and supporting studies illustrates both the need for automation and the difficulty of establishing behavioral equivalence. Pontolillo and Mousavi manually produced comparable Qiskit, Cirq, and Q\# implementations for a multilingual testing benchmark and documented language differences encountered during translation~\cite{Pontolillo2022}; Su\'arez et al.\ derived a taxonomy of Qiskit migration scenarios and used it to guide LLM-assisted refactoring of code for newer APIs~\cite{Suarez2025}; and Cardinal et al.\ showed that a Qiskit version migration could preserve circuits, optimizers, and Hamiltonians yet substantially alter results because of a hidden change in sampling behavior~\cite{Cardinal2026}. More directly, Regan et al.\ automatically translated a restricted subset of Qiskit gate-model programs to an adiabatic representation~\cite{Regan2019}; Hans and Groppe developed a source-to-source translator from a subset of Silq to Qiskit~\cite{Hans2022}; and Wesley's LinguaQuanta translates between OpenQASM and Quipper while addressing challenges such as ancilla management and round-trip stability~\cite{Wesley2024}. Recent work has explored LLM-based translation from Qiskit to Cirq, although its reported evaluation was limited to a small circuit example~\cite{Siavash2025}, and rule-based translation of dynamic OpenQASM~3 programs to CUDA-Q C++ kernels~\cite{Kulkarni2026}.

\section{Methods}
\label{sec:methods}

Here, we describe the methods by which this work was performed, including the autonomous migration, ensemble-validation pipeline, and the subsequent generation and cross-model evaluation of reusable migration skills. ~\autoref{fig:pipeline} summarizes the overall experimental workflow.
 
\subsection{Benchmark Suite and Starting Point}
\label{subsec:qedc}

For the QED-C Application-Oriented Performance Benchmarks~\cite{QEDC2019,Lubinski2023}, we used Git revision \texttt{2e7a983} of the public suite together with local compatibility updates for CUDA-Q~0.14. At the start of the study, the suite already included CUDA-Q implementations for Hidden Shift, Bernstein--Vazirani Method~1, Quantum Fourier Transform Method~1, Phase Estimation, and HamLib Hamiltonian Simulation Methods~1--3~\cite{Brown2026}. These implementations, particularly Hidden Shift, served as structural examples for the new ports. They provided examples of the benchmark execution interface, result analysis plumbing, and CUDA-Q kernel organization. The migration set comprised thirteen methods spanning a variety of quantum programming paradigms and techniques as summarized in Table~\ref{tab:qedc-benchmarks}.


\begin{table*}[!t]
\centering
\footnotesize
\setlength{\tabcolsep}{4pt}
\renewcommand{\arraystretch}{0.96}
\caption{QED-C benchmarks included in the migration study. Mid-circuit measurement indicates whether a benchmark requires measurement during circuit execution.}
\label{tab:qedc-benchmarks}
\begin{tabularx}{\textwidth}{@{}>{\raggedright\arraybackslash}p{0.18\textwidth}
                  >{\raggedright\arraybackslash}X
                  >{\centering\arraybackslash}p{0.075\textwidth}@{}}
\toprule
\textbf{Benchmark} & \textbf{Description} & \textbf{Mid-circuit meas.} \\
\midrule
Bernstein--Vazirani Method 2 &
Determines a hidden bitstring $s$ of length $n$ from a black-box function
$f(x)=s\cdot x \bmod 2$, demonstrating linear quantum speedup. Method 2
uses only two qubits plus mid-circuit measurements to emulate the $n$-qubit
register, mathematically equivalent to method 1 but with a much smaller
physical width. & Yes \\

QFT Method 2 &
Initializes the qubits to represent the Fourier-basis encoding of a classical
value, then performs the inverse QFT to retrieve the original value. Tests that
$\mathrm{QFT}^{-1}\circ\mathrm{QFT}$ round-trips correctly; correct output is a
single state with 100\% probability. & No \\

QFT Method 3 &
Initializes the qubits to a quantum superposition, performs an inverse QFT, and
compares the resulting distribution against the analytically expected one.
Designed to exercise QFT on an entangled state---currently marked as a
work-in-progress because depolarizing errors are difficult to detect with this
construction. & No \\

Grovers &
Quantum unstructured-search algorithm that finds a single ``correct'' item
among $N$ candidates in $\sim\sqrt{N}$ oracle calls, a quadratic speedup over
the classical $O(N)$. The benchmark fixes one marked item per circuit and
compares measured counts to the analytical Grover-amplified distribution. & No \\

Deutsch--Jozsa &
Decides whether a black-box Boolean function is constant (returns all 0s or all
1s) or balanced (returns 0 on exactly half its inputs) in a single oracle
query---the first quantum algorithm to demonstrate an exponential separation
from classical query complexity. The benchmark runs both a constant and a
balanced oracle at each width. & No \\

Amplitude Estimation &
Estimates the amplitude $a$ of a ``good state'' in a unitary preparation
$A\lvert 0\rangle=\sqrt{1-a}\lvert\psi_0\rangle\lvert 0\rangle+
\sqrt{a}\lvert\psi_1\rangle\lvert 1\rangle$ using phase estimation on Grover's
amplification operator, achieving a quadratic speedup over classical Monte
Carlo sampling. Random $a=\sin^2(n/2^k)$ values are chosen so the answer lands
exactly on the counting-register grid. & No \\

Monte Carlo Method 1 &
Estimates $\mathbb{E}[f(X)]$ for a random variable $X$ via amplitude
estimation, with method 1 using arbitrary distributions and functions
(default: truncated Gaussian with variable mean and $f(x)=x^2$). Quadratic
Monte Carlo speedup, but circuits are deep because of the general-purpose
state-preparation and function-evaluation operators. & No \\

Monte Carlo Method 2 &
Uses the same expected-value estimation goal as method 1, but with an extremely
simple distribution (uniform via Hadamards) and a balanced-oracle-style
function. Resulting circuits are much shallower than method 1, useful for
evaluating the amplitude-estimation core without the state-preparation
overhead. & No \\

MaxCut Method 1 &
Solves the NP-hard MaxCut optimization problem on an undirected graph using the
QAOA ansatz. Method 1 simply executes the ansatz circuit at fixed angles at
different widths to gauge per-system circuit-execution fidelity, without the
optimization loop. & No \\

MaxCut Method 2 &
Runs the complete MaxCut/QAOA algorithm: the ansatz circuit is executed many
times under a classical optimizer (default COBYLA) that searches for the angles
maximizing the cut approximation ratio. Reports per-iteration metrics for the
variational loop. & No \\

HHL &
Solves the linear system $Ax=b$ for a sparse Hermitian $A$ (at most $s$
nonzeros per row or column) and prepares
$\lvert x\rangle\propto A^{-1}\lvert b\rangle$, combining QPE, Hamiltonian
simulation, and a uniformly controlled rotation for eigenvalue inversion.
Fidelity is computed against the classical solution to $Ax=b$. & Yes \\

Shor's Method 1 &
Order-finding subroutine of Shor's factoring algorithm---the standard textbook
construction with three quantum registers and $4n+2$ qubits, doing the full QPE
on the modular-exponentiation operator. & No \\

Shor's Method 2 &
Uses the same order-finding goal as method 1, but reduces the counting register
from $2n$ qubits to a single qubit by using mid-circuit measurement and
classical feedforward to play the role of the $2n$ control qubits iteratively.
Same algorithmic output, smaller circuit width. & Yes \\
\bottomrule
\end{tabularx}
\end{table*}

\subsection{Models and Agents}
\label{subsec:models}

In this work, we study autonomous quantum-software migration using general-purpose models rather than models trained specifically for a particular quantum language or algorithm. Three general-purpose coding agents performed the migration independently in separate working trees:

\begin{itemize}
\item Anthropic Claude Opus~4.7 (Extra-High) through Claude Code~2.1.156~\cite{Anthropic2026a}.
\item OpenAI GPT~5.5 (Extra-High) through Codex~0.131.0~\cite{OpenAI2026a}.
\item Google Gemini~3.1 Pro through Cursor Agent CLI (2026.05.28-a70ca7c)~\cite{Cursor2026a,Cursor2026b}.
\end{itemize}

Due to the probabilistic / chaotic nature in agentic coding, the anonymized names Models~1--3 are used throughout the results to discourage comparison based on a single porting effort per model. Cross contamination, where agents read history, skills, code, etc. from other ports, is a significant risk due to the variety of locations where information is stored. We therefore separated ports from each model such that each started with a clean context (no agent history whatsoever) with OS-level permission controls preventing non-root access to history and work by another agent.

\subsection{Migration Protocol}
\label{subsec:protocol}

Each agent received the same initial prompt and was instructed to port one benchmark method at a time from the Qiskit implementation without changing the higher-level quantum algorithm. The agents were directed to the CUDA-Q documentation, the \texttt{cuda-q-academic} training repository, and the existing \texttt{hidden\_shift} port as a template. The agents could inspect and modify files, execute shell commands, consult the supplied documentation, and run the benchmark tests. A Python file for running the benchmarks using the Qiskit backend was supplied and the agent was instructed to add a CUDA-Q test for each benchmark as the ports progressed. A method advanced to the next stage after its noiseless automated test met the configured fidelity floor. Once the initial port was complete, with all CUDA-Q benchmarks passing the agent test script, the ports were evaluated with a more extensive test suite exercising alternative benchmark options and sweeping through a range of circuit widths. Follow-up instructions needed to resolve identified issues were counted as additional user prompts.

Validation was performed using NVIDIA CUDA-Q~0.14~\cite{Kim2023} with the \texttt{nvidia} statevector backend~\cite{Bayraktar2023}. Reference executions used Qiskit~2.4.1 and Qiskit Aer~0.17.2~\cite{JavadiAbhari2024}. The evaluation used noiseless simulation rather than quantum hardware so that migration behavior could be separated from device noise.

The initial prompt was:

\begin{tcolorbox}[
  enhanced,
  breakable,
  title={Initial Prompt},
  colback=gray!5,
  colframe=black!65,
  colbacktitle=black!8,
  coltitle=black,
  fonttitle=\bfseries,
  boxrule=0.5pt,
  arc=1mm,
  left=6pt,right=6pt,top=5pt,bottom=5pt,
  before skip=8pt,after skip=10pt]
\small
We are going to add CUDA-Q support to many of the benchmarks. Some of the benchmarks have options for multiple methods and we will do one method at a time. The existing framework in \texttt{hidden\_shift} should be used as an example. Do the ports to CUDA-Q from the qiskit version. Do not change the higher level (quantum) algorithm used in qiskit. To be effective you will need to understand the documentation for qiskit and CUDA-Q. CUDA-Q academic provides training and examples using CUDA-Q and you should use the repo here: \url{https://github.com/NVIDIA/cuda-q-academic}. The documentation and Python API for CUDA-Q are here: \url{https://nvidia.github.io/cuda-quantum/0.14.0/}, and the tests in the CUDA-Q 0.14.2 repo might also be useful. For each benchmark/method we add, we will enable the test case in the untracked \texttt{test\_benchmarks.py} in this directory. For each benchmark, write a separate \texttt{.md} file in the ``learnings'' directory that contains implementation notes and the number of code-line changes required. Additionally, write a separate file for the benchmark in the skills directory that will have information to ultimately be consolidated to create a skill for you to be more efficient in implementing or porting to CUDA-Q. The learnings and skills directories should remain untracked. Let's start by running the test script and making sure the currently implemented benchmarks pass.
\end{tcolorbox}

After the baseline ports were complete, each agent was given all three implementations of each benchmark and access to the agent generated ``learnings'' files. Each was asked to select a preferred implementation. The assessment considered correctness, fidelity to the source algorithm, scalability, robustness, and implementation quality. The resulting votes and rationales were then used to consolidate complementary features into a single implementation rather than assuming that one agent would be best for every method.

\subsection{Evaluation Measures}
\label{subsec:metrics}

Migration effort was characterized by (i) the number of user prompts required to reach the automated fidelity threshold, (ii) agent turns, comprising assistant steps and tool calls, and (iii) tokens consumed. Turn and token accounting is not necessarily identical across agent harnesses. We therefore emphasize per-method shares of each model's total rather than directly comparing raw turn or token counts across harnesses.

Implementation size was measured as the percentage change in framework-specific source lines relative to the Qiskit implementation, excluding comments. Methods implemented in a shared file are reported at the algorithm-family level. Circuit structure was measured as the percentage change in total and two-qubit gate counts relative to Qiskit at two stages: the algorithmic representation and the circuit after Qiskit transpilation or CUDA-Q simulation lowering. These counts characterize differences in representation and lowering; they are not direct measures of execution time or hardware performance. Because CUDA-Q and Qiskit represent composite and multi-controlled operations at different abstraction levels, gate-count results were interpreted together with fidelity and runtime behavior rather than as standalone correctness measures.

The cross-agent assessment provided a complementary qualitative check on algorithm fidelity and implementation quality. Because a benchmark's test and analyzer can share assumptions with its implementation, passing a fidelity threshold was not taken to imply that independently produced ports were semantically equivalent.

\subsection{Skill Generation and Evaluation}
\label{subsec:skills}

The open Agent Skills specification defines a skill as a portable package of instructions and metadata that an agent can load when relevant to a task, optionally accompanied by executable scripts, reference material, templates, and other resources~\cite{AgentSkills2026}. Skills augment the context and procedures available to an agent without modifying the parameters of its underlying LLM. In this study, we developed a skill as a structured text artifact that captures reusable, task-specific procedures, implementation patterns, constraints, common failure modes, and validated workarounds. The intent was to generate reusable guidance to improve agent success and code quality while reducing token usage and agent turns.

Beyond the working code, each porting session produced a per-benchmark “skill” document distilling the specific CUDA-Q learnings. For each model, two skills were generated -- a test skill and a final skill. The test skill was generated by the agent after porting the first eleven methods, through HHL in Table~\ref{tab:qedc-benchmarks}, before any exposure to porting Shor's Methods~1 and~2. These were deliberately withheld so that the skills could be evaluated on unseen algorithms in clean agent contexts. Following the completion of Shor's, a final skill for each model was generated to be consolidated for public release.

For assessment of the test skills, each model ported the Shor's methods under four conditions: without a skill and with the skill produced by each of Models~1--3. Because the skill-evaluation runs were performed in clean contexts, they were separate from the initial Shor's ports that were performed as the last two benchmarks in the main porting effort. The latter was used for migration effort and cross-agent assessment; the former was used only for skill assessment. Skill-assisted runs were used only for the skill-efficiency analysis. Benefit was calculated as the percentage reduction in turns and tokens relative to the corresponding model-and-method ``no-skill'' baseline. Applying every skill to every model measured both self-use and transfer to a different model. Finally, the reusable material was consolidated and divided into separate references for general CUDA-Q algorithm implementation and code-migration-specific guidance.

\FloatBarrier
\section{Results}
\label{sec:results}

\subsection{Migration Effort}
\label{subsec:effort}

All initial ports were completed without direct coding assistance aside from help to Model~3 to use \texttt{run()} instead of \texttt{sample()} for mid-circuit measurement and guidance to Model~1 that assumptions made about JIT caching were incorrect. Subsequent testing with the larger test suite identified a number of issues, however, and required additional prompts to achieve a satisfactory port. Some of the issues involved shortcuts taken when the agents encountered significant challenges these could be egregious. Model~3 lowered the fidelity threshold for a benchmark in the test script down to 50\% accuracy and declared the port a success. Model~1 announced the invention of ``qiskit-as-a-dsl'' with a ``port'' that ran mostly using the original language. In both cases, the agents resorted to shortcuts after multiple hours of failure in honest attempts. All three models violated the initial prompt with changes to the high-level algorithm such as increasing the number of qubits used to eliminate mid-circuit measurement. Models 1 and 2 introduced unacceptable restrictions such as hard-coding kernels for the specific qubit count used in the test or removing features not tested in the initial script. These two models also resorted to using a low-level ``builder mode'' rather than the CUDA-Q abstraction for troublesome benchmarks. Interestingly, direct help was not required in resolving the issues other than instructing the agent why the result was unacceptable and that it needed to be fixed.

Table~\ref{tab:user-prompts} reports the user prompts required for each baseline port. Grover's, Deutsch--Jozsa, and amplitude estimation required one prompt from every model. The larger prompt counts were concentrated in a few model--method combinations: Monte Carlo Method~1 required six prompts for Model~1 but one for Model~3, whereas HHL required seven prompts for Model~3 but two for Model~2. Model~2 required 20 prompts across all thirteen methods, compared with 26 for Model~3 and 33 for Model~1.


\begin{table}[t]
\centering
\caption{User prompts required to obtain a satisfactory port. Blue identifies the lowest value in each row and red the highest; tied values share the corresponding highlight.}
\label{tab:user-prompts}
\begin{tabular}{lrrr}
\toprule
Benchmark & Model 1 & Model 2 & Model 3 \\
\midrule
BV Method 2 & \best{1} & \best{1} & \worst{3} \\
QFT Method 2 & \best{1} & \worst{2} & \worst{2} \\
QFT Method 3 & \worst{2} & \best{1} & \best{1} \\
Grovers & \best{1} & \best{1} & \best{1} \\
Deutsch--Jozsa & \best{1} & \best{1} & \best{1} \\
Amplitude Estimation & \best{1} & \best{1} & \best{1} \\
Monte Carlo Method 1 & \worst{6} & 3 & \best{1} \\
Monte Carlo Method 2 & \worst{3} & \best{1} & \best{1} \\
MaxCut Method 1 & \worst{3} & 2 & \best{1} \\
MaxCut Method 2 & \worst{2} & \best{1} & \best{1} \\
HHL & 4 & \best{2} & \worst{7} \\
Shor's Method 1 & \worst{3} & \best{1} & \best{1} \\
Shor's Method 2 & \worst{3} & \best{1} & \worst{3} \\
\midrule
\textbf{Total} & \textbf{33} & \textbf{20} & \textbf{26} \\
\bottomrule
\end{tabular}
\end{table}

The turn and token shares in Tables~\ref{tab:turn-share} and~\ref{tab:token-share} show that effort was unevenly distributed within each run. For Model~1, HHL and Monte Carlo Method~1 together accounted for 40.13\% of turns and 45.61\% of tokens. HHL alone accounted for 48.44\% of Model~3's turns and 51.31\% of its tokens, making it the clearest single outlier. This was driven by repeated iteration on the sparse Hamiltonian simulation subroutine and the mid-circuit ancilla measurement plumbing. For Monte Carlo Method~1, Model~1's effort was largely spent adjusting the encoding of dynamic controlled-RY sequences that come from an unknown-arity polynomial expansion. Model~2's effort was more broadly distributed: Monte Carlo Method~1 was its largest token share at 17.45\%, followed by HHL at 16.61\% and MaxCut Method~1 at 12.74\%. Shor's Method~2 was also unusually costly for Model~3, using 16.35\% of its tokens, although its turn share was 9.47\%. 


\begin{table}[t]
\centering
\caption{Percentage of each model's total turns attributable to each task. Blue identifies the lowest share in each row and red the highest.}
\label{tab:turn-share}
\begin{tabular}{lrrr}
\toprule
Task & Model 1 & Model 2 & Model 3 \\
\midrule
Initial Prompt & \best{0.64\%} & 1.03\% & \worst{1.47\%} \\
BV Method 2 & \worst{7.48\%} & 5.31\% & \best{4.69\%} \\
QFT Method 2 & \best{1.85\%} & \worst{8.05\%} & 2.39\% \\
QFT Method 3 & \worst{3.42\%} & 2.40\% & \best{1.29\%} \\
Grovers & 3.71\% & \worst{4.28\%} & \best{1.75\%} \\
Deutsch--Jozsa & 2.64\% & \worst{4.62\%} & \best{1.10\%} \\
Amplitude Estimation & 3.78\% & \worst{6.51\%} & \best{2.11\%} \\
Monte Carlo Method 1 & \worst{19.89\%} & 15.07\% & \best{5.51\%} \\
Monte Carlo Method 2 & 4.70\% & \best{2.57\%} & \worst{9.56\%} \\
MaxCut Method 1 & 8.13\% & \worst{14.55\%} & \best{4.14\%} \\
MaxCut Method 2 & \worst{4.99\%} & 4.62\% & \best{2.11\%} \\
HHL & 20.24\% & \best{13.70\%} & \worst{48.44\%} \\
Shor's Method 1 & \worst{8.41\%} & 7.02\% & \best{3.58\%} \\
Shor's Method 2 & 6.49\% & \best{4.28\%} & \worst{9.47\%} \\
\bottomrule
\end{tabular}

\medskip
\noindent Raw totals: 1,384 turns for Model 1, 584 for Model 2, and 1,088 for Model 3.
\end{table}


\begin{table}[t]
\centering
\caption{Percentage of each model's total tokens attributable to each task. Blue identifies the lowest share in each row and red the highest.}
\label{tab:token-share}
\begin{tabular}{lrrr}
\toprule
Task & Model 1 & Model 2 & Model 3 \\
\midrule
Initial Prompt & 0.12\% & \worst{0.51\%} & \best{0.11\%} \\
BV Method 2 & \best{2.30\%} & \worst{5.32\%} & 4.73\% \\
QFT Method 2 & \best{0.75\%} & \worst{9.58\%} & 3.49\% \\
QFT Method 3 & 2.23\% & \worst{3.44\%} & \best{0.26\%} \\
Grovers & 1.79\% & \worst{6.26\%} & \best{0.42\%} \\
Deutsch--Jozsa & 1.43\% & \worst{2.69\%} & \best{0.30\%} \\
Amplitude Estimation & 2.33\% & \worst{5.11\%} & \best{0.68\%} \\
Monte Carlo Method 1 & \worst{20.16\%} & 17.45\% & \best{2.22\%} \\
Monte Carlo Method 2 & 5.31\% & \best{3.75\%} & \worst{5.37\%} \\
MaxCut Method 1 & 9.98\% & \worst{12.74\%} & \best{2.78\%} \\
MaxCut Method 2 & \worst{6.16\%} & 4.11\% & \best{1.50\%} \\
HHL & 25.45\% & \best{16.61\%} & \worst{51.31\%} \\
Shor's Method 1 & \worst{10.28\%} & \best{4.02\%} & 5.47\% \\
Shor's Method 2 & 8.87\% & \best{3.07\%} & \worst{16.35\%} \\
\bottomrule
\end{tabular}

\medskip
\noindent Raw totals: 637 thousand tokens for Model 1, 74 thousand for Model 2, and 199 thousand for Model 3.
\end{table}

Within each model, prompt count closely tracked the distribution of work: across the thirteen methods, its Pearson correlation with turn share ranged from 0.85 to 0.92 and with token share from 0.87 to 0.95. Thus, despite differences in harness accounting, repeated user intervention was a useful within-run indicator of difficult ports. The generally lower cost of later methods in the QFT, Monte Carlo, and MaxCut families also indicates that reusable kernel structures amortized part of the cost after the first method established a workable pattern.

\subsection{Implementation Size and Circuit Structure}
\label{subsec:structure}

Table~\ref{tab:framework-code-lines} shows the change in framework-specific lines of code relative to the Qiskit reference. Framework-independent helpers were factored out; only the CUDA-Q-specific paths are counted here.
For most benchmarks, the CUDA-Q port is shorter than the Qiskit source---up to 73\% shorter for MaxCut (Model~1) and 56\% shorter for HHL (Model~1)---because CUDA-Q's decorator-mode kernels and native multi-controlled gate forms encode nested Qiskit constructions more concisely. Two benchmarks break the trend. Monte Carlo grew by 40\%--133\% across all three models, because the dynamic gate sequences must be encoded as parallel arrays walked by a runtime dispatcher inside a kernel rather than emitted directly onto a \texttt{QuantumCircuit}; Model~2's variant was longest because it additionally maintains a swap-based control-reordering table. Shor's grew by 55\% for Model~2 because of the size of its per-op-kind unrolled walker; Model~1 shrank Shor's by 22\% by encoding gate records in a compact class-based emitter. HHL grew by 44\% for Model~3 for two reasons: the port retains Qiskit's module structure across two files, and it uses \texttt{cudaq.adjoint()} inside the kernel, which is inelegant but did not require the manual inverse constructions that Model~1 and Model~2 wrote.


\begin{table}[t]
\centering
\caption{Percentage change in lines of framework-specific code relative to the Qiskit implementation, excluding comments. Blue marks the largest reduction in each row and red the largest increase.}
\label{tab:framework-code-lines}
\begin{tabular}{lrrr}
\toprule
Benchmark family & Model 1 & Model 2 & Model 3 \\
\midrule
BV & \worst{9.84\%} & \best{$-6.56$\%} & 3.28\% \\
QFT & $-32.67$\% & \worst{$-19.33$\%} & $-31.33$\% \\
Grovers & $-46.30$\% & \best{$-50.93$\%} & $-46.30$\% \\
Deutsch--Jozsa & $-10.77$\% & \best{$-20.00$\%} & $-10.77$\% \\
Amplitude Estimation & \best{$-17.17$\%} & \worst{16.16\%} & 3.03\% \\
Monte Carlo & 42.37\% & \worst{133.33\%} & \best{40.11\%} \\
MaxCut & \best{$-72.51$\%} & $-23.17$\% & \worst{$-2.11$\%} \\
HHL & \best{$-55.93$\%} & $-40.65$\% & \worst{44.07\%} \\
Shor's & \best{$-21.97$\%} & \worst{54.75\%} & 12.13\% \\
\bottomrule
\end{tabular}
\end{table}

Tables~\ref{tab:two-qubit-gate-count} and~\ref{tab:total-gate-count} report gate counts at both the algorithmic (untranspiled, native-multi-controlled) and transpiled-simulation (fixed one/two-qubit basis) levels. Most of the algorithmic differences between the three models are within one percent, because the ports share the same underlying quantum algorithm and the same primitive dispatch. The dramatic reductions relative to Qiskit come from the native multi-controlled gate forms in CUDA-Q's algorithmic layer: Amplitude Estimation drops by 79\%--83\% in total algorithmic gates, Monte Carlo Method~2 by 96\%, and Monte Carlo Method~1 by 83\%, in every case because Qiskit expresses the same operator as a decomposition into \texttt{cx} and \texttt{u3} while CUDA-Q counts each multi-controlled gate once. 

Grover's is an interesting counter-example: the algorithmic two-qubit count is 600\% larger than Qiskit across all three ports, but the transpiled two-qubit count is 96\% smaller. In this case, Qiskit's algorithmic counter records the multi-controlled X as a single gate while the CUDA-Q counter attributes each control as a two-qubit contribution. At the transpiled level the Qiskit toolchain fully decomposes the multi-controlled X into two-qubit primitives and the CUDA-Q simulator does not. All three models converge on the same reductions to within a fraction of a percent for almost every benchmark, because the transpilation target and the simulator basis are the same regardless of the porting model. Deutsch--Jozsa is the only case where transpiled-simulation total-gate counts diverge non-trivially between models: Model~1 emits zero simulator-visible gates (the constant oracle is elided entirely as a no-op), while Model~2 and Model~3 retain a symbolic \texttt{constant\_oracle} kernel that the simulator still visits.


\begin{table*}[!t]
\centering
\caption{Percentage change in two-qubit gate count relative to Qiskit. Blue marks the largest reduction and red the largest increase in each row; all omitted HHL transpiled/simulation values were unavailable in the source.}
\label{tab:two-qubit-gate-count}
\begin{minipage}{0.48\textwidth}
\centering
\textbf{(a) Algorithmic}

\begin{tabular}{lrrr}
\toprule
Benchmark & Model 1 & Model 2 & Model 3 \\
\midrule
BV Method 2 & 0.00\% & 0.00\% & 0.00\% \\
QFT Method 2 & 0.00\% & 0.00\% & 0.00\% \\
QFT Method 3 & 0.00\% & 0.00\% & 0.00\% \\
Grovers & \worst{600.00\%} & \worst{600.00\%} & \worst{600.00\%} \\
Deutsch--Jozsa & 0.00\% & 0.00\% & 0.00\% \\
Amplitude Estimation & \best{$-57.02$\%} & $-45.75$\% & $-45.75$\% \\
Monte Carlo Method 1 & $-55.13$\% & $-55.13$\% & $-55.13$\% \\
Monte Carlo Method 2 & $-88.14$\% & \best{$-89.25$\%} & $-88.14$\% \\
MaxCut Method 1 & 0.00\% & 0.00\% & 0.00\% \\
MaxCut Method 2 & 0.00\% & 0.00\% & 0.00\% \\
HHL & 16.44\% & 16.44\% & 16.44\% \\
Shor's Method 1 & 4.42\% & 4.42\% & 4.42\% \\
Shor's Method 2 & 7.41\% & 7.41\% & 7.41\% \\
\bottomrule
\end{tabular}
\end{minipage}
\hfill
\begin{minipage}{0.48\textwidth}
\centering
\textbf{(b) Transpiled/simulation}

\begin{tabular}{lrrr}
\toprule
Benchmark & Model 1 & Model 2 & Model 3 \\
\midrule
BV Method 2 & 0.00\% & 0.00\% & 0.00\% \\
QFT Method 2 & $-50.00$\% & $-50.00$\% & $-50.00$\% \\
QFT Method 3 & $-50.00$\% & $-50.00$\% & $-50.00$\% \\
Grovers & $-96.11$\% & $-96.11$\% & $-96.11$\% \\
Deutsch--Jozsa & \best{$-100.00$\%} & \worst{0.00\%} & \worst{0.00\%} \\
Amplitude Estimation & \best{$-16.19$\%} & \worst{3.24\%} & \worst{3.24\%} \\
Monte Carlo Method 1 & $-26.31$\% & $-26.31$\% & $-26.31$\% \\
Monte Carlo Method 2 & $-84.61$\% & \best{$-85.99$\%} & \worst{$-83.24$\%} \\
MaxCut Method 1 & 0.00\% & 0.00\% & 0.00\% \\
MaxCut Method 2 & 0.00\% & 0.00\% & 0.00\% \\
HHL & --- & --- & --- \\
Shor's Method 1 & $-54.13$\% & $-54.13$\% & $-54.13$\% \\
Shor's Method 2 & $-51.00$\% & $-51.00$\% & $-51.00$\% \\
\bottomrule
\end{tabular}
\end{minipage}
\end{table*}


\begin{table*}[!t]
\centering
\caption{Percentage change in total gate count relative to Qiskit. Blue marks the largest reduction and red the largest increase in each row; all omitted HHL transpiled/simulation values were unavailable in the source.}
\label{tab:total-gate-count}
\begin{minipage}{0.48\textwidth}
\centering
\textbf{(a) Algorithmic}

\begin{tabular}{lrrr}
\toprule
Benchmark & Model 1 & Model 2 & Model 3 \\
\midrule
BV Method 2 & 0.00\% & 0.00\% & 0.00\% \\
QFT Method 2 & 0.00\% & 0.00\% & 0.00\% \\
QFT Method 3 & 0.00\% & 0.00\% & 0.00\% \\
Grovers & 1.04\% & 1.04\% & 1.04\% \\
Deutsch--Jozsa & \worst{39.36\%} & \best{0.00\%} & \best{0.00\%} \\
Amplitude Estimation & \best{$-83.42$\%} & $-78.79$\% & \worst{$-78.75$\%} \\
Monte Carlo Method 1 & $-82.83$\% & $-82.83$\% & \worst{$-82.81$\%} \\
Monte Carlo Method 2 & \best{$-96.74$\%} & $-96.31$\% & \worst{$-96.30$\%} \\
MaxCut Method 1 & 0.00\% & 0.00\% & 0.00\% \\
MaxCut Method 2 & 0.00\% & 0.00\% & 0.00\% \\
HHL & $-0.09$\% & $-0.09$\% & $-0.09$\% \\
Shor's Method 1 & $-26.45$\% & $-26.45$\% & \worst{$-26.25$\%} \\
Shor's Method 2 & $-22.88$\% & \best{$-23.07$\%} & \worst{$-22.78$\%} \\
\bottomrule
\end{tabular}
\end{minipage}
\hfill
\begin{minipage}{0.48\textwidth}
\centering
\textbf{(b) Transpiled/simulation}

\begin{tabular}{lrrr}
\toprule
Benchmark & Model 1 & Model 2 & Model 3 \\
\midrule
BV Method 2 & $-41.61$\% & $-41.61$\% & $-41.61$\% \\
QFT Method 2 & $-67.08$\% & $-67.08$\% & $-67.08$\% \\
QFT Method 3 & $-68.58$\% & $-68.58$\% & $-68.58$\% \\
Grovers & $-94.40$\% & $-94.40$\% & $-94.40$\% \\
Deutsch--Jozsa & \best{$-100.00$\%} & \worst{$-44.74$\%} & \worst{$-44.74$\%} \\
Amplitude Estimation & \best{$-75.91$\%} & $-69.19$\% & \worst{$-69.11$\%} \\
Monte Carlo Method 1 & \best{$-84.49$\%} & \best{$-84.49$\%} & \worst{$-84.47$\%} \\
Monte Carlo Method 2 & \best{$-95.45$\%} & $-94.85$\% & \worst{$-94.84$\%} \\
MaxCut Method 1 & $-13.33$\% & $-13.33$\% & $-13.33$\% \\
MaxCut Method 2 & $-13.33$\% & $-13.33$\% & $-13.33$\% \\
HHL & --- & --- & --- \\
Shor's Method 1 & \best{$-73.34$\%} & \best{$-73.34$\%} & \worst{$-73.27$\%} \\
Shor's Method 2 & \best{$-72.92$\%} & $-72.86$\% & \worst{$-72.91$\%} \\
\bottomrule
\end{tabular}
\end{minipage}
\end{table*}

\subsection{Cross-Agent Assessment}
\label{subsec:voting}

Table~\ref{tab:best-implementation-votes} records each model's preferred implementation after comparing all three baseline ports. Model~2 received a majority for seven of thirteen methods, Model~1 for four, and Model~3 for one; Grover's had no majority. No implementation source was therefore preferred across the full suite.

The rationales reveal complementary strengths. Model~1 was preferred for robust handling of CUDA-Q-specific edge cases, including qubit reset in Bernstein--Vazirani Method~2, avoidance of a latent loop-bound issue in QFT Method~3, and arbitrary-arity controlled operations in Monte Carlo Method~1. Model~2 was preferred where method isolation or parity with the Qiskit benchmark was decisive, including QFT Method~2, both MaxCut methods, HHL, and both Shor's methods. Model~3 received the amplitude-estimation majority because it exposed relevant state parameters, although its inverse-QFT implementation had an issue with \texttt{rz.ctrl} versus \texttt{r1.ctrl} use.

The Grover assessments exposed a severe endianness bug in Model~3 that the existing implementation-specific test path had not detected. This case shows why the ensemble assessment added information beyond a pass/fail fidelity result. Shor's Method~2 had an elegant implementation by Model~2, but Model~1 developed a more scalable per-bit feedback decomposition ($O(n^2)$ vs a $O(2^n)$ integer-equality fan-out). For QFT Method~3, Model~3 wrote a two-line ``\texttt{safe\_secret\_int}'' bounds clamp that prevents the MLIR compiler from unrolling loops beyond the \texttt{qvector} allocation size; this fix earned Model~3 votes not only for Method~3 but also unblocked JIT lowering of the shared kernel's other methods.

The cross-agent assessment was fruitful in identifying implementation differences, latent bugs, and preferred directions, with all models providing a critical assessment of their own work. The full 3-model assessment was used for agent-driven consolidation. The consolidated benchmark code was not restricted to the work of a single model but instead combined the best choices at a finer granularity. This was used to generate the final code for testing and publication.


\begin{table*}[!t]
\centering
\caption{Each model's vote for the best implementation. Orange, green, and blue identify votes for Models 1, 2, and 3, respectively.}
\label{tab:best-implementation-votes}
\begin{tabularx}{\textwidth}{l l l l X}
\toprule
Benchmark & Model 1 vote & Model 2 vote & Model 3 vote & Rationale \\
\midrule
BV Method 2 & \voteone & \voteone & \votethree & Better qubit reset \\
QFT Method 2 & \votetwo & \votetwo & \votetwo & Avoids compiling unrelated method code \\
QFT Method 3 & \votetwo & \votetwo & \votethree & Model 3 chosen for the best MLIR loop-bound workaround; latent bug with Model 1 \\
Grovers & Equal & \votetwo & \voteone & Style, correct gate-count statistics, and a severe bug with Model 3 \\
Deutsch--Jozsa & \voteone & \voteone & \voteone & Elegant branch removal \\
Amplitude Estimation & \voteone & \votethree & \votethree & Limited state-qubit support in Model 2; Model 3 maintains \texttt{psi\_zero}/\texttt{psi\_one} as arguments but issue with \texttt{rz.ctrl} versus \texttt{r1.ctrl} \\
Monte Carlo Method 1 & \voteone & \voteone & \voteone & Handles MCRY/MCX of any arity without swap shuffling and with lower circuit depth \\
Monte Carlo Method 2 & \votetwo & \voteone & \voteone & Style; same \texttt{rz.ctrl} divergence in IQFT for Model 3 \\
MaxCut Method 1 & \votetwo & \votetwo & \voteone & Better parity with Qiskit \\
MaxCut Method 2 & \votetwo & \votetwo & \voteone & Better parity with Qiskit \\
HHL & \votetwo & \voteone & \votetwo & Model 1 uses list encoding elegantly; Model 3 maintains \texttt{cudaq.adjoint} \\
Shor's Method 1 & \votetwo & \votetwo & \votetwo & Style \\
Shor's Method 2 & \voteone & \votetwo & \votetwo & Model 1 for scalability; Model 2 avoids unnecessary machinery \\
\bottomrule
\end{tabularx}
\end{table*}

\subsection{Skill Assessment}
\label{subsec:skill-results}

Tables~\ref{tab:skill-turn-reduction} and~\ref{tab:skill-token-reduction} measure how much effort each model saves when supplied with a skill authored by itself or by a different model. Skills were evaluated on porting both algorithmic variants of Shor's in a clean context. This differs from the previous results with Shor's where the port was performed with the full history of the eleven other ports. Every non-zero entry represents a strict improvement over the no-skill baseline, meaning skills are always beneficial regardless of authorship. The magnitude of the benefit is substantial: turn reductions of 31\%--85\% and token reductions of 5\%--96\%, with a median well above 50\%.

Two patterns stand out. First, cross-model skill transfer works: the largest single reduction in both tables (Model~3 porting under a Model~1 skill: 85\% fewer turns, 96\% fewer tokens) is a cross-model pairing, and the two smallest reductions (Model~2 porting under a Model~3 skill and Model~1 porting under a Model~2 skill for one method) are also cross-model but still yield meaningful savings. Therefore, the best skill for a given porting model is not, in general, the one written by that same model. The cross-model gains indicate that the skills captured transferable CUDA-Q knowledge rather than merely reproducing the originating model's preferred workflow. Second, Model~3 gains the most from any skill in both turns and tokens; Model~2 gains the least, consistent with its already low baseline effort. Aggregated across method families, the average token reduction with the best-available skill is 67\%, indicating that the marginal cost of a distilled skill (produced once) is repaid many times over across subsequent ports, and that this repayment is largely insensitive to which model authored the skill.

Common themes in the skills included guidance on endianness, mid-circuit measurement, control-of-adjoint composition, and gate mapping. The length of a skill can impact token counts, and we made no attempt to optimize skill length in these experiments. For the consolidated skill now published with CUDA-Q, we have divided components helpful for general implementation from those specific to migration with the intent of further reducing token usage.


\begin{table*}[!t]
\centering
\caption{Percentage reduction in model turns when a skill generated by one model was supplied to another model. Rows identify the porting model and columns the skill source; blue marks the largest reduction and red the smallest within each row.}
\label{tab:skill-turn-reduction}
\begin{tabular}{clrrrrr}
\toprule
Method & Porting model & No skill & Model 1 skill & Model 2 skill & Model 3 skill & Maximum \\
\midrule
1 & Model 1 & \worst{0.00\%} & 50.00\% & 25.00\% & \best{60.71\%} & \textbf{60.71\%} \\
1 & Model 2 & \worst{0.00\%} & \best{68.24\%} & 54.12\% & 48.24\% & \textbf{68.24\%} \\
1 & Model 3 & \worst{0.00\%} & \best{60.00\%} & 38.46\% & 56.92\% & \textbf{60.00\%} \\
2 & Model 1 & \worst{0.00\%} & 31.25\% & 6.25\% & \best{37.50\%} & \textbf{37.50\%} \\
2 & Model 2 & \worst{0.00\%} & 13.79\% & \best{31.03\%} & 3.45\% & \textbf{31.03\%} \\
2 & Model 3 & \worst{0.00\%} & \best{84.62\%} & 51.28\% & 80.34\% & \textbf{84.62\%} \\
\bottomrule
\end{tabular}
\end{table*}


\begin{table*}[!t]
\centering
\caption{Percentage reduction in token count when a skill generated by one model was supplied to another model. Rows identify the porting model and columns the skill source; blue marks the largest reduction and red the smallest within each row.}
\label{tab:skill-token-reduction}
\begin{tabular}{clrrrrr}
\toprule
Method & Porting model & No skill & Model 1 skill & Model 2 skill & Model 3 skill & Maximum \\
\midrule
1 & Model 1 & \worst{0.00\%} & 5.24\% & 28.74\% & \best{53.30\%} & \textbf{53.30\%} \\
1 & Model 2 & \worst{0.00\%} & 30.38\% & 30.63\% & \best{63.71\%} & \textbf{63.71\%} \\
1 & Model 3 & \worst{0.00\%} & \best{76.31\%} & 63.17\% & 51.69\% & \textbf{76.31\%} \\
2 & Model 1 & \worst{0.00\%} & 15.44\% & \best{45.73\%} & 8.30\% & \textbf{45.73\%} \\
2 & Model 2 & \worst{0.00\%} & \best{84.67\%} & 63.58\% & 74.01\% & \textbf{84.67\%} \\
2 & Model 3 & \worst{0.00\%} & \best{95.92\%} & 75.79\% & 92.41\% & \textbf{95.92\%} \\
\bottomrule
\end{tabular}
\end{table*}

\section{Discussion}
\label{sec:discussion}

\subsection{Autonomous Migration}
\label{subsec:limitations}

This case study shows that current general-purpose autonomous coding agents can complete a substantial quantum-framework migration when they are given executable tests, examples in the target framework, and permission to iterate. All three systems produced working CUDA-Q ports for a diverse set of QED-C benchmark methods, including algorithms with mid-circuit measurement, controlled arithmetic, inverse QFTs, variational optimization, and nontrivial classical post-processing. In our experience, the general models were not advanced to the point where one can expect high-quality one-shot ports where more sophisticated algorithms are involved. Nonetheless, the entire effort for each model required on the order of several business days for a satisfactory port -- a drastic improvement over the weeks or months we estimate would be required for a human engineer without agent assistance.

The study also motivates continued research into specialized agents with models tuned for quantum programming and validation to further reduce the human effort required. In addition to substantial differences in the intuition and concepts central to quantum algorithms, significant semantic differences between languages can contribute to challenges. Relevant to this case, Qiskit exposes a circuit object with indexed quantum and classical registers, composition, inversion, controlled subcircuits, parameter binding, and decompositions. CUDA-Q expresses the same algorithms most naturally as Python-decorated kernels with explicit qubit operations, different result objects, and stricter kernel-language constraints.

\subsection{Where Migration Became Difficult}
\label{subsec:difficulty}

Difficulty was driven less by source length than by mismatches between programming models. The expensive trajectories were associated with features for which Qiskit and CUDA-Q expose different abstractions: mid-circuit measurement and classical feed-forward, dynamic multi-controlled rotations, inverse and controlled kernel composition, parameter handling during compilation, and modular arithmetic. HHL combined several of these issues and became the dominant outlier for Model~3. Monte Carlo Method~1 likewise dominated Model~1's effort while requiring substantially more framework-specific code in every port. In contrast, later methods that could reuse an established family-level structure were generally less expensive. This pattern suggests that, for autonomous migration, the first correct solution to a framework mismatch is a reusable asset whose value extends beyond the individual benchmark.

The gate-count results reinforce this interpretation. Large changes were frequently shared across all three agents and sometimes reversed between the algorithmic and lowered representations. Such agreement points to differences in framework primitives and compiler boundaries rather than agent quality. Gate totals should therefore be treated as structural diagnostics unless they are paired with a common target, compilation policy, and performance measurement.

\subsection{Value of an Agent Ensemble}
\label{subsec:ensemble}

The cross-agent assessments show that no single optimization criterion identified the preferred port. Model~1 was often favored for avoiding CUDA-Q-specific failure modes and for scalable control logic; Model~2 was often favored for isolation, simplicity, or feature parity; and Model~3 offered elegant solutions for parameterization with amplitude estimation and for avoiding JIT unrolling pitfalls. Consolidation could therefore combine strengths that were distributed across independently generated implementations.

Given the shortcuts taken during the initial porting process, violation of the prompt restrictions, and the survival of latent bugs through initial testing, cross-agent assessment would seem a valuable validation tool as resource limitations most often preclude full testing coverage of every possibility. Agreement with a single numerical threshold (fidelity in this case) is therefore necessary but insufficient for migration: verification should also compare algorithm semantics, output conventions, and supported parameter ranges. Independently implemented alternatives proved to be useful in identifying areas for enhancement in code generated by each model. Circuit-structure measures, here limited to total and two-qubit gate counts, provide a useful validation signal but differences might not always indicate an algorithmic change. In Grover's search and Monte Carlo, for example, CUDA-Q native controlled operations and Qiskit decompositions were counted at different levels of abstraction.

\subsection{Reusable Skills}
\label{subsec:skill-discussion}

The clean-context Shor's experiments provide evidence that knowledge accumulated during the other eleven migrations was reusable on a new algorithm family. All eighteen skill-assisted model--method pairings reduced both measured turns and tokens. More importantly, the strongest skill often originated from another model. This result supports treating agent-generated guidance as shareable project infrastructure: implementation constraints and effective patterns can transfer across model families even when their unaided trajectories differ.

\subsection{Practical Implications}
\label{subsec:implications}

Prompt engineering refers to the deliberate design and refinement of instructions, context, examples, constraints, and output requirements used to elicit desired model behavior~\cite{White2023}. Prior work has shown that model performance can vary with semantically equivalent changes in wording and formatting, and that effective prompts may not transfer reliably across models or tasks~\cite{Leidinger2023,Sclar2024}. A controlled evaluation of prompt engineering would therefore require systematically varied prompt formulations and repeated trials for each model--agent combination. Such an evaluation is outside the scope of this case study, which held the initial prompt constant across agents and treated subsequent interactions as part of the observed migration workflow. The initial prompt was concise relative to the scope of the migration and informally structured, leaving several behavioral constraints and edge cases implicit. Although this limits conclusions about the best performance attainable through prompt optimization, it reflects a plausible practitioner interaction: exhaustively anticipating every implementation contingency before migration begins is rarely economical or feasible, particularly for users who are domain experts rather than specialists in coding-agent design. Within this setting, the study supports several practical recommendations. 

We can summarize our key learnings and suggestions as follows. The agent should be given a test script together with an explicit instruction that the script must not be modified. Post-validation using alternative tests not visible to the agent can identify over-specialization to the original test script. The prompt should establish clear rules for the extent to which it can change algorithms, parameter support, etc. and also rules on use of the programming model (e.g., to prohibit CUDA-Q builder mode, assembly language, etc.).  Expect that the agent can violate the prompt. Output metrics such as sampling probability and expectation value can be insufficient for validation on their own.  Supplementing with measures of circuit structure and cross-model assessment might help reduce issues identified on manual review. In testing, instruct the agent to test without support for using the original language (in this case, run the CUDA-Q tests without Qiskit installed). Port methods sequentially and document newly identified issues in a skill to reduce turns and token usage. 

\section{Summary}
\label{sec:conclusion}

This study demonstrates both the power and the current limitations of agentic coding for quantum software. Three general-purpose coding agents completed a substantial migration of the QED-C benchmark suite from Qiskit to CUDA-Q with little direct coding assistance and on a timescale of days, demonstrating a level of practical capability that would have been difficult to anticipate only a short time ago. As a result, we have been able to publish CUDA-Q implementations for the QED-C Application-Oriented Benchmarks. 

The agent failures were equally informative: difficult mismatches between programming models led agents to alter algorithms, restrict functionality, and exploit weaknesses in incomplete tests. These results do not yet support unattended migration, but they show that agentic coding, when coupled with executable validation and independent assessment, can compress a demanding quantum-software engineering effort from weeks or months to days. Our results demonstrate that agent skills are an effective approach for documenting reusable knowledge to improve future performance, and we have released a skill that captures the CUDA-Q implementation learnings with the intent of improving user experiences with agentic coding and migration.

Together, the released implementations and reusable skill demonstrate that agentic coding is becoming a practical force multiplier for quantum computing---one capable of broadening framework support, retaining specialized knowledge, and accelerating the path from quantum algorithm to robust software.

\section*{Code Availability}

The quantum application programs described in this work are implemented in the QED-C open-source Application-Oriented Benchmark suite, available at \url{https://github.com/SRI-International/QC-App-Oriented-Benchmarks}. The \texttt{cudaq-multi-agent-staged} branch currently holds the new implementations as they are merged into main. The CUDA-Q skills are available at \url{https://github.com/NVIDIA/skills/tree/main/skills} and also with CUDA-Q at \url{https://github.com/NVIDIA/cuda-quantum}.

\section*{Acknowledgments}

AC was supported in part by the U.S. Department of Energy, Office of Science, National Quantum Information Science Research Centers, Quantum Systems Accelerator (Award No. DE-SCL0000121), under Contract No. DE-AC02-05CH11231, and partly by the National Energy Research Scientific Computing Center (NERSC), a U.S. Department of Energy Office of Science User Facility (Project No. m888).



\bibliographystyle{ieeetr}
\bibliography{references}

@misc{AgentSkills2026,
  author       = {{Agent Skills Project}},
  title        = {Agent Skills Specification},
  year         = {2026},
  howpublished = {\url{https://agentskills.io/specification}},
  note         = {Accessed September 30, 2026}
}

@misc{Anthropic2026a,
  author       = {{Anthropic}},
  title        = {{Claude Code}: Command-Line Interface Reference},
  year         = {2026},
  howpublished = {Claude Code Documentation},
  url          = {https://code.claude.com/docs/en/cli-usage},
  note         = {Accessed September 3, 2026}
}

@misc{Anthropic2026b,
  author       = {{Anthropic}},
  title        = {{Claude Opus 4.7 System Card}},
  year         = {2026},
  month        = apr,
  howpublished = {\url{https://www.anthropic.com/system-cards}},
  note         = {Accessed October 8, 2026}
}

@inproceedings{Basit2025,
  author    = {Basit, Abdul and Shao, Minghao and Asif, Muhammad Haider and Innan, Nouhaila and Kashif, Muhammad and Marchisio, Alberto and Shafique, Muhammad},
  title     = {{QHackBench}: Benchmarking Large Language Models for Quantum Code Generation Using {PennyLane} Hackathon Challenges},
  booktitle = {2025 IEEE International Conference on Quantum Artificial Intelligence},
  pages     = {316--322},
  year      = {2025},
  doi       = {10.1109/QAI63978.2025.00056},
  url       = {https://arxiv.org/abs/2506.20008}
}

@article{Bayraktar2023,
  author        = {Bayraktar, Harun and Charara, Ali and Clark, David and Cohen, Saul and Costa, Timothy and Fang, Yao-Lung L. and Gao, Yang and Guan, Jack and Gunnels, John and Haidar, Azzam and Hehn, Andreas and Hohnerbach, Markus and Jones, Matthew and Lubowe, Tom and Lyakh, Dmitry and Morino, Shinya and Springer, Paul and Stanwyck, Sam and Terentyev, Igor and Varadhan, Satya and Wong, Jonathan and Yamaguchi, Takuma},
  title         = {{cuQuantum SDK}: A High-Performance Library for Accelerating Quantum Science},
  journal       = {arXiv},
  year          = {2023},
  eprint        = {2308.01999},
  archivePrefix = {arXiv},
  primaryClass  = {quant-ph},
  doi           = {10.48550/arXiv.2308.01999},
  url           = {https://arxiv.org/abs/2308.01999}
}

@article{Bergholm2018,
  author        = {Bergholm, Ville and Izaac, Josh and Schuld, Maria and Gogolin, Christian and others},
  title         = {{PennyLane}: Automatic Differentiation of Hybrid Quantum-Classical Computations},
  journal       = {arXiv},
  year          = {2018},
  eprint        = {1811.04968},
  archivePrefix = {arXiv},
  primaryClass  = {quant-ph},
  doi           = {10.48550/arXiv.1811.04968},
  url           = {https://arxiv.org/abs/1811.04968}
}

@article{Brown2026,
  author  = {Brown, W. Michael and Ramesh, Anurag and Lubinski, Thomas and Nguyen, Thien and {Bernal Neira}, David E.},
  title   = {{Multi-GPU} Quantum Circuit Simulation and the Impact of Network Performance},
  journal = {Computer Physics Communications},
  volume  = {324},
  pages   = {110126},
  year    = {2026},
  month   = jul,
  issn    = {0010-4655},
  doi     = {10.1016/j.cpc.2026.110126},
  url     = {https://doi.org/10.1016/j.cpc.2026.110126}
}

@inproceedings{Campbell2025,
  author    = {Campbell, Charlie and Chen, Hao Mark and Luk, Wayne and Fan, Hongxiang},
  title     = {Enhancing {LLM}-based Quantum Code Generation with Multi-Agent Optimization and Quantum Error Correction},
  booktitle = {Proceedings of the 62nd ACM/IEEE Design Automation Conference},
  pages     = {1--7},
  year      = {2025},
  doi       = {10.1109/DAC63849.2025.11133316},
  url       = {https://arxiv.org/abs/2504.14557}
}

@inproceedings{Cardinal2026,
  author    = {Cardinal, Julien and Benzarti, Imen and El Boussaidi, Ghizlane and Pere, Christophe},
  title     = {Migrating {QAOA} from {Qiskit} 1.x to 2.x: An Experience Report},
  booktitle = {Proceedings of the 7th IEEE/ACM International Workshop on Quantum Software Engineering},
  year      = {2026},
  doi       = {10.1145/3786150.3788612},
  url       = {https://doi.org/10.1145/3786150.3788612}
}

@article{Cassano2023,
  author  = {Cassano, Federico and Gouwar, John and Nguyen, Daniel and Nguyen, Sydney and Phipps-Costin, Luna and Pinckney, Donald and Yee, Ming-Ho and Zi, Yangtian and Anderson, Carolyn Jane and Feldman, Molly Q. and Guha, Arjun and Greenberg, Michael and Jangda, Abhinav},
  title   = {{MultiPL-E}: A Scalable and Polyglot Approach to Benchmarking Neural Code Generation},
  journal = {IEEE Transactions on Software Engineering},
  volume  = {49},
  number  = {7},
  pages   = {3675--3691},
  year    = {2023},
  doi     = {10.1109/TSE.2023.3267446},
  url     = {https://doi.org/10.1109/TSE.2023.3267446}
}

@article{Cassano2024,
  author  = {Cassano, Federico and Gouwar, John and Lucchetti, Francesca and Schlesinger, Claire and Freeman, Anders and Anderson, Carolyn Jane and Feldman, Molly Q. and Greenberg, Michael and Jangda, Abhinav and Guha, Arjun},
  title   = {Knowledge Transfer from High-Resource to Low-Resource Programming Languages for Code {LLM}s},
  journal = {Proceedings of the ACM on Programming Languages},
  volume  = {8},
  number  = {OOPSLA2},
  pages   = {677--708},
  year    = {2024},
  doi     = {10.1145/3689735},
  url     = {https://doi.org/10.1145/3689735}
}

@book{Cirq2025,
  author    = {{Cirq Developers}},
  title     = {{Cirq}: Python Package for Writing, Manipulating, and Running Quantum Circuits on Quantum Computers and Simulators},
  publisher = {Zenodo},
  year      = {2025},
  doi       = {10.5281/zenodo.4062499},
  url       = {https://doi.org/10.5281/zenodo.4062499}
}

@misc{Cursor2026a,
  author       = {{Cursor}},
  title        = {{Cursor CLI}},
  year         = {2026},
  howpublished = {Cursor Documentation},
  url          = {https://cursor.com/docs/cli/overview},
  note         = {Accessed September 3, 2026}
}

@misc{Cursor2026b,
  author       = {{Cursor}},
  title        = {Models and Pricing},
  year         = {2026},
  howpublished = {Cursor Documentation},
  url          = {https://cursor.com/docs/models-and-pricing},
  note         = {Accessed September 3, 2026}
}

@inproceedings{Dupuis2024,
  author    = {Dupuis, Nicolas and Buratti, Luca and Vishwakarma, Sanjay and Viudes Forrat, Aitana and Kremer, David and Faro, Ismael and Puri, Ruchir and Cruz-Benito, Juan},
  title     = {{Qiskit Code Assistant}: Training {LLM}s for Generating Quantum Computing Code},
  booktitle = {2024 IEEE LLM Aided Design Workshop},
  pages     = {1--4},
  year      = {2024},
  doi       = {10.1109/LAD62341.2024.10691762},
  url       = {https://arxiv.org/abs/2405.19495}
}

@misc{Google2026,
  author       = {{Google DeepMind}},
  title        = {{Gemini 3.1 Pro Model Card}},
  year         = {2026},
  month        = feb,
  howpublished = {\url{https://deepmind.google/models/model-cards/gemini-3-1-pro/}},
  note         = {Accessed October 8, 2026}
}

@inproceedings{Guo2024,
  author    = {Guo, Xiaoyu and Zhao, Jianjun and Zhao, Pengzhan},
  title     = {On Repairing Quantum Programs Using {ChatGPT}},
  booktitle = {Proceedings of the 5th ACM/IEEE International Workshop on Quantum Software Engineering},
  pages     = {9--16},
  year      = {2024},
  doi       = {10.1145/3643667.3648223},
  url       = {https://arxiv.org/abs/2401.14913}
}

@inproceedings{Guo2025,
  author    = {Guo, Xiaoyu and Wang, Minggu and Zhao, Jianjun},
  title     = {{QuanBench}: Benchmarking Quantum Code Generation with Large Language Models},
  booktitle = {Proceedings of the 40th IEEE/ACM International Conference on Automated Software Engineering},
  pages     = {2657--2669},
  year      = {2025},
  doi       = {10.1109/ASE63991.2025.00218},
  url       = {https://arxiv.org/abs/2510.16779}
}

@inproceedings{Hans2022,
  author    = {Hans, Julian and Groppe, Sven},
  title     = {{Silq2Qiskit}---Developing a Quantum Language Source-to-Source Translator},
  booktitle = {Proceedings of the 6th International Conference on Computer Science and Software Engineering},
  pages     = {86--95},
  year      = {2022},
  doi       = {10.1145/3569966.3570114},
  url       = {https://doi.org/10.1145/3569966.3570114}
}

@article{Heim2020,
  author  = {Heim, Bettina and Soeken, Mathias and Marshall, Sarah and Granade, Christopher E. and Roetteler, Martin and Geller, Alan and Troyer, Matthias and Svore, Krysta M.},
  title   = {Quantum Programming Languages},
  journal = {Nature Reviews Physics},
  volume  = {2},
  pages   = {709--722},
  year    = {2020},
  doi     = {10.1038/s42254-020-00245-7},
  url     = {https://doi.org/10.1038/s42254-020-00245-7}
}

@article{Henderson2025,
  author  = {Henderson, Elena R. and Henderson, Jessie M. and Ange, Joshua and Thornton, Mitchell A.},
  title   = {Programming quantum computers with large language models},
  journal = {arXiv},
  year    = {2025},
  eprint  = {2506.18125},
  archivePrefix = {arXiv},
  doi     = {10.48550/arXiv.2506.18125},
  url     = {https://arxiv.org/abs/2506.18125}
}

@article{JavadiAbhari2024,
  author        = {Javadi-Abhari, Ali and Treinish, Matthew and Krsulich, Kevin and Wood, Christopher J. and Lishman, Jake and Gacon, Julien and Martiel, Simon and Nation, Paul D. and Bishop, Lev S. and Cross, Andrew W. and others},
  title         = {Quantum Computing with {Qiskit}},
  journal       = {arXiv},
  year          = {2024},
  eprint        = {2405.08810},
  archivePrefix = {arXiv},
  primaryClass  = {quant-ph},
  doi           = {10.48550/arXiv.2405.08810},
  url           = {https://arxiv.org/abs/2405.08810}
}

@inproceedings{Jern2025,
  author    = {Jern, Linus and Uotila, Valter and Yu, Cong and Zhao, Bo},
  title     = {{Agent-Q}: Fine-Tuning Large Language Models for Quantum Circuit Generation and Optimization},
  booktitle = {2025 IEEE International Conference on Quantum Computing and Engineering},
  pages     = {1621--1632},
  year      = {2025},
  doi       = {10.1109/QCE65121.2025.00179},
  url       = {https://zbjob.github.io/QCE25-AgentQ.pdf}
}

@article{Jimenez2023,
  title={Swe-bench: Can language models resolve real-world github issues?, 2024},
  author={Jimenez, Carlos E and Yang, John and Wettig, Alexander and Yao, Shunyu and Pei, Kexin and Press, Ofir and Narasimhan, Karthik},
  journal={URL https://arxiv. org/abs/2310.06770},
  volume={7},
  year={2023}
}

@inproceedings{Kim2023,
  author    = {Kim, Jin-Sung and McCaskey, Alex and Heim, Bettina and Modani, Manish and Stanwyck, Sam and Costa, Timothy B.},
  title     = {{CUDA Quantum}: The Platform for Integrated Quantum-Classical Computing},
  booktitle = {Proceedings of the 60th ACM/IEEE Design Automation Conference},
  pages     = {1--4},
  year      = {2023},
  doi       = {10.1109/DAC56929.2023.10247886},
  url       = {https://doi.org/10.1109/DAC56929.2023.10247886}
}

@article{Kocetkov2023,
  author  = {Kocetkov, Denis and Li, Raymond and Ben Allal, Loubna and Li, Jia and Mou, Chenghao and Mu{\~n}oz Ferrandis, Carlos and Jernite, Yacine and Mitchell, Margaret and Hughes, Sean and Wolf, Thomas and Bahdanau, Dzmitry and von Werra, Leandro and de Vries, Harm},
  title   = {The Stack: 3 {TB} of Permissively Licensed Source Code},
  journal = {Transactions on Machine Learning Research},
  year    = {2023},
  url     = {https://arxiv.org/abs/2211.15533}
}

@article{Kulkarni2026,
  author        = {Kulkarni, Vinooth and Lee, Jaehyun and Hutchings, Adam and Albahri, Anas and Nana, Jai and Xu, Shuai and Chaudhary, Vipin},
  title         = {Efficient Transpilation of {OpenQASM} 3.0 Dynamic Circuits to {CUDA-Q}: Performance and Expressiveness Advantages},
  journal       = {arXiv},
  year          = {2026},
  eprint        = {2604.11599},
  archivePrefix = {arXiv},
  primaryClass  = {quant-ph},
  doi           = {10.48550/arXiv.2604.11599},
  url           = {https://arxiv.org/abs/2604.11599}
}

@inproceedings{Leidinger2023,
  author    = {Leidinger, Alina and van Rooij, Robert and Shutova,
               Ekaterina},
  title     = {The Language of Prompting: What Linguistic Properties
               Make a Prompt Successful?},
  booktitle = {Findings of the Association for Computational
               Linguistics: EMNLP 2023},
  year      = {2023},
  pages     = {9210--9232},
  address   = {Singapore},
  publisher = {Association for Computational Linguistics},
  doi       = {10.18653/v1/2023.findings-emnlp.618},
  url       = {https://aclanthology.org/2023.findings-emnlp.618/}
}

@ARTICLE{Lubinski2023,
  author={Lubinski, Thomas and Johri, Sonika and Varosy, Paul and Coleman, Jeremiah and Zhao, Luning and Necaise, Jason and Baldwin, Charles H. and Mayer, Karl and Proctor, Timothy},
  journal={IEEE Transactions on Quantum Engineering}, 
  title={Application-Oriented Performance Benchmarks for Quantum Computing}, 
  year={2023},
  volume={4},
  number={},
  pages={1-32},
  doi={10.1109/TQE.2023.3253761}
}

@article{Lubinski2024,
  title={Quantum Algorithm Exploration using Application-Oriented Performance Benchmarks},
  author={Lubinski, Thomas and Goings, Joshua J and Mayer, Karl and Johri, Sonika and Reddy, Nithin and Mehta, Aman and Bhatia, Niranjan and Rappaport, Sonny and Mills, Daniel and Baldwin, Charles H and others},
  journal={arXiv preprint arXiv:2402.08985},
  year={2024}
}

@inproceedings{Mikuriya2025,
  author    = {Mikuriya, Taku and Ishigaki, Tatsuya and Kawarada, Masayuki and Minami, Shunya and Kadowaki, Tadashi and Suzuki, Yohichi and Naito, Soshun and Takata, Shunya and Kato, Takumi and Basseda, Tamotsu and Yamada, Reo and Takamura, Hiroya},
  title     = {{QCoder Benchmark}: Bridging Language Generation and Quantum Hardware through Simulator-Based Feedback},
  booktitle = {Proceedings of the 18th International Natural Language Generation Conference},
  pages     = {743--752},
  year      = {2025},
  publisher = {Association for Computational Linguistics},
  url       = {https://aclanthology.org/2025.inlg-main.43/}
}

@misc{NVIDIA2026a,
  author       = {{NVIDIA}},
  title        = {{CUDA-Q Language Specification}},
  howpublished = {NVIDIA CUDA-Q Documentation},
  year         = {2026},
  note         = {[Online]. Available: https://nvidia.github.io/cuda-quantum/latest/specification/cudaq.html. Accessed: Oct. 3, 2026}
}

@misc{NVIDIA2026b,
  author       = {{NVIDIA Corporation}},
  title        = {{cuQuantum SDK}: Getting Started---Framework Integrations},
  year         = {2026},
  howpublished = {NVIDIA cuQuantum Documentation},
  url          = {https://docs.nvidia.com/cuda/cuquantum/latest/getting-started/index.html},
  note         = {Accessed September 3, 2026}
}

@misc{OpenAI2026a,
  author       = {{OpenAI}},
  title        = {{Codex CLI}},
  year         = {2026},
  howpublished = {OpenAI Documentation},
  url          = {https://learn.chatgpt.com/docs/codex/cli},
  note         = {Accessed September 3, 2026}
}

@misc{OpenAI2026b,
  author       = {{OpenAI}},
  title        = {{GPT-5.5 System Card}},
  year         = {2026},
  month        = apr,
  howpublished = {\url{https://openai.com/index/gpt-5-5-system-card/}},
  note         = {Accessed October 8, 2026}
}

@inproceedings{Pan2024,
  author    = {Pan, Rangeet and Ibrahimzada, Ali Reza and Krishna, Rahul and Sankar, Divya and Wassi, Lambert Pouguem and Merler, Michele and Sobolev, Boris and Pavuluri, Raju and Sinha, Saurabh and Jabbarvand, Reyhaneh},
  title     = {Lost in Translation: A Study of Bugs Introduced by Large Language Models While Translating Code},
  booktitle = {Proceedings of the 46th IEEE/ACM International Conference on Software Engineering},
  pages     = {995--1007},
  year      = {2024},
  publisher = {ACM},
  doi       = {10.1145/3597503.3639226},
  url       = {https://doi.org/10.1145/3597503.3639226}
}

@inproceedings{Pontolillo2022,
  author    = {Pontolillo, Gabriel and Mousavi, Mohammad Reza},
  title     = {A Multi-Lingual Benchmark for Property-Based Testing of Quantum Programs},
  booktitle = {Proceedings of the 3rd International Workshop on Quantum Software Engineering},
  pages     = {1--7},
  year      = {2022},
  doi       = {10.1145/3528230.3528395},
  url       = {https://doi.org/10.1145/3528230.3528395}
}

@misc{QEDC2019,
    author = {{QED-C}},
    title = {{Application-Oriented Benchmarks for Quantum Computing}},
    year = {2019},
    url = {https://github.com/SRI-International/QC-App-Oriented-Benchmarks}
}

@inproceedings{Regan2019,
  author    = {Regan, Malcolm and Eastwood, Brody and Nagabhiru, Mahita and Mueller, Frank},
  title     = {Automatically Translating Quantum Programs from a Subset of Common Gates to an Adiabatic Representation},
  booktitle = {Reversible Computation},
  series    = {Lecture Notes in Computer Science},
  volume    = {11497},
  pages     = {146--161},
  year      = {2019},
  publisher = {Springer},
  url       = {https://arcb.csc.ncsu.edu/~mueller/ftp/pub/mueller/papers/rc19-2.pdf}
}

@inproceedings{Sclar2024,
  author    = {Sclar, Melanie and Choi, Yejin and Tsvetkov, Yulia
               and Suhr, Alane},
  title     = {Quantifying Language Models' Sensitivity to Spurious
               Features in Prompt Design or: How I Learned to Start
               Worrying about Prompt Formatting},
  booktitle = {The Twelfth International Conference on Learning
               Representations},
  series    = {ICLR 2024},
  year      = {2024},
  url       = {https://proceedings.iclr.cc/paper_files/paper/2024/hash/6c0e99d736da621403018ca7b32b1a4d-Abstract-Conference.html}
}

@inproceedings{Siavash2025,
  author    = {Siavash, Nazanin and Moin, Armin},
  title     = {{LLM}-Powered Quantum Code Transpilation},
  booktitle = {2025 IEEE International Conference on Quantum Computing and Engineering},
  pages     = {410--411},
  year      = {2025},
  doi       = {10.1109/QCE65121.2025.10370},
  url       = {https://arxiv.org/abs/2507.12480}
}

@article{Slim2026,
  author  = {Slim, Ali and Hamieh, Haydar and Kotaich, Jawad and Ghosn, Yehya and Chehimi, Mahdi and Mohanna, Ammar and Hammoud, Hasan Abed Al Kader and Ghanem, Bernard},
  title   = {{QuanBench+}: A unified multi-framework benchmark for {LLM}-based quantum code generation},
  journal = {arXiv},
  year    = {2026},
  eprint  = {2604.08570},
  archivePrefix = {arXiv},
  doi     = {10.48550/arXiv.2604.08570},
  url     = {https://arxiv.org/abs/2604.08570}
}

@article{Suarez2025,
  author        = {Su{\'a}rez, Jos{\'e} Manuel and Bibb{\'o}, Luis Mariano and Bogado, Joaquin and Fernandez, Alejandro},
  title         = {Automatic Qiskit Code Refactoring Using Large Language Models},
  journal       = {arXiv},
  year          = {2025},
  eprint        = {2506.14535},
  archivePrefix = {arXiv},
  primaryClass  = {cs.SE},
  doi           = {10.48550/arXiv.2506.14535},
  url           = {https://arxiv.org/abs/2506.14535}
}

@article{Vishwakarma2024,
  author  = {Vishwakarma, Sanjay and Harkins, Francis and Golecha, Siddharth and Bajpe, Vishal Sharathchandra and Dupuis, Nicolas and Buratti, Luca and Kremer, David and Faro, Ismael and Puri, Ruchir and Cruz-Benito, Juan},
  title   = {{Qiskit HumanEval}: An evaluation benchmark for quantum code generative models},
  journal = {arXiv},
  year    = {2024},
  eprint  = {2406.14712},
  archivePrefix = {arXiv},
  doi     = {10.48550/arXiv.2406.14712},
  url     = {https://arxiv.org/abs/2406.14712}
}

@article{Wang2023,
  title={Voyager: An open-ended embodied agent with large language models},
  author={Wang, Guanzhi and Xie, Yuqi and Jiang, Yunfan and Mandlekar, Ajay and Xiao, Chaowei and Zhu, Yuke and Fan, Linxi and Anandkumar, Anima},
  journal={arXiv preprint arXiv:2305.16291},
  year={2023}
}

@inproceedings{Wesley2024,
  author    = {Wesley, Scott},
  title     = {{LinguaQuanta}: Towards a Quantum Transpiler Between {OpenQASM} and {Quipper}},
  booktitle = {Reversible Computation},
  pages     = {142--160},
  year      = {2024},
  publisher = {Springer},
  doi       = {10.1007/978-3-031-62076-8_10},
  url       = {https://arxiv.org/abs/2404.08147}
}

@inproceedings{White2023,
  author    = {White, Jules and Fu, Quchen and Hays, Sam and Sandborn,
               Michael and Olea, Carlos and Gilbert, Henry and Elnashar,
               Ashraf and Spencer-Smith, Jesse and Schmidt, Douglas C.},
  title     = {A Prompt Pattern Catalog to Enhance Prompt Engineering
               with {ChatGPT}},
  booktitle = {Proceedings of the 30th Conference on Pattern Languages
               of Programs},
  series    = {PLoP 2023},
  year      = {2023},
  publisher = {Association for Computing Machinery},
  doi       = {10.64346/PLoP2023p05},
  url       = {https://doi.org/10.64346/PLoP2023p05}
}

@article{Yang2024,
  title={Swe-agent: Agent-computer interfaces enable automated software engineering},
  author={Yang, John and Jimenez, Carlos and Wettig, Alexander and Lieret, Kilian and Yao, Shunyu and Narasimhan, Karthik and Press, Ofir},
  journal={Advances in Neural Information Processing Systems},
  volume={37},
  pages={50528--50652},
  year={2024}
}

@article{Zhang2025,
  author  = {Zhang, Hanliang and David, Cristina and Wang, Meng and Paulsen, Brandon and Kroening, Daniel},
  title   = {Scalable, Validated Code Translation of Entire Projects Using Large Language Models},
  journal = {Proceedings of the ACM on Programming Languages},
  volume  = {9},
  number  = {PLDI},
  pages   = {1616--1641},
  year    = {2025},
  doi     = {10.1145/3729315},
  url     = {https://doi.org/10.1145/3729315}
}

\clearpage

\end{document}